\documentclass[10pt,showkeys,pra,aps]{revtex4-1}
\pdfoutput=1
\usepackage{amssymb}
\usepackage{amsmath}
\usepackage{graphicx}
\usepackage{dcolumn}
\usepackage{bm}
\usepackage[caption=false]{subfig}
\usepackage{siunitx}

\begin{document}

\title{Thermal Conductivity Measurement of Supported Thin Film Materials Using the 3$\omega$ Method }
\author{Daxi Zhang}
\author{Amir Behbahanian}
\author{Nicholas A. Roberts}%
\email{nick.roberts@usu.edu}
\affiliation{ 
Department of Mechanical and Aerospace Engineering, Utah State University, Logan, UT
}%

\date{\today}

\begin{abstract}
{\it In this article, we are proposing a thorough analysis of the cross, and the in-plane thermal conductivity of thin-film materials based on the 3$\omega$ method. The analysis accommodates a 2D mathematical heat transfer model of a semi-infinite body and the details of the sample preparation followed by the measurement process. The presented mathematical model for the system considers a two-dimensional space for its solution. It enables the calculation of the cross-plane thermal conductivity with a single frequency measurement, the derived equation opens new opportunities for frequency-based and penetration-depth dependent thermal conductivity analysis. The derived equation for the in-plane thermal conductivity is dependent on the cross-plane thermal conductivity. Both in and cross-plane thermal conductivities enable the measurements in two steps of measurements, the resistance-temperature slope measurement and another set of measures that extracts the third harmonic of the voltage signal. We evaluated the methodology in two sets of samples, silicon nitride and boron nitride, both on silicon wafers.  We observed anisotropic thermal conductivity in the cross and the in-plane direction, despite the isotropic nature of the thin films, which we relate to the total anisotropy of the thin film-substrate system. The technique is conducive to the thermal analysis of next-generation nanoelectronic devices.  }
\end{abstract}

\maketitle

\section*{INTRODUCTION}

The 3$\omega$ method was developed in the early 1990s by D. G. Cahill\cite{wang2009analysis}. This method is valuable for measuring the thermal properties of various systems, specifically, thin films\cite{hu20063,yusibani2009application,schiffres2011improved,lee2013temperature}.  Compared to other measurement methods, the 3$\omega$ method has many advantages. For example, the 3$\omega$ method reduces the effect of radiation at the low temperature region \cite{cahill1990thermal}. It is also remarkably faster than any other traditional steady-state measurement techniques, since the 3$\omega$ method belongs to the transient measurement technique category\cite{hanninen2013implementing}. The equipment setup for the 3$\omega$ method is relatively inexpensive\cite{koh2009comparison}. Additionally the 3$\omega$ method is one of the most accurate thermal property measurement methods for thin film materials, especially for isotropic thin film materials\cite{dames2013measuring}. However in industrial applications, there are many insulating thin film materials that are anisotropic\cite{jannot2011measurement}. A new technique needs to be developed for measuring anisotropic thin films, in which the thermal conductivity depends on direction.

A variety of techniques have been developed to determine the cross-plane thermal conductivity. Only few have been focused on measuring the in-plane thermal conductivity of thin film materials. For example, Bonger et. al\cite{bogner2017cross} developed an in-plane thermal properties measurement that requires an additional series of multiple heaters of different widths. This method showed a significant difference between the theoretical results and the experimental results. Mavrokefalos et. al\cite{mavrokefalos2007four} further developed a suspended film structure to create two suspended platforms, but involves more complicated fabrication procedures. Mishra et al.\cite{mishra20153} developed a technique for the direct calculation of the thermal conductivity tensor of anisotropic materials. Specifically they performed measurements for the c-axis and ab-axis on mica samples. They also performed measurements on machined and polished mica at multiple angles, which agreed well with the computed thermal conductivity values from the tensor. However, the
probes for measuring the thermal conductivity tensor were deposited on a mica sample with thickness of 0.25”. It is difficult to use same technique on anisotropic thin film materials.
		
We report an improved 3$\omega$ method for measuring the in-plane thermal conductivity of supported thin film structures based on the concept of the 3$\omega$ method. Comparing with method developed by Mishra et al., we are not focused on measurements for the full thermal conductivity tensor, but are focused on anisotropic ratio of the in-plane and cross-plane thermal conductivities and can thus utilize a more simplified setup. Comparing with the suspended film method, the fabrication procedure is simpler. Moreover, this method is capable of reducing uncertainty by repeating the measurement at different locations. The derivation of the mathematical model for the method, the fabrication procedure, and the experimental results of the 100nm pre-grown LPCVD silicon nitride film on silicon wafer, and the 64nm sputter deposited amorphous boron nitride thin film on silicon wafer, are presented.
\section{DEVELOPMENT OF THE MATHEMATICAL MODEL FOR IN-PLANE THERMAL CONDUCTIVITY MEASUREMENT}
In the traditional 3$\omega$ method, a narrow metal line is deposited on the top of the film of interest. This metal line is used as a heater as well as a temperature probe. Because the 3$\omega$ method involves generating periodic heating with an AC current source (in general it is more common to use a voltage source with few limitations, see ref.\cite{dames20051}), the film of interest is typically covered with the metal deposited directly on top. If the film is not dielectric, a thin isolation layer should be used between the film and the metal electrodes. The essential idea for determining the thermal properties of the thin film is by measuring the temperature change of the metal line. The temperature change of the metal line is approximately equal to the temperature change at the surface of the film. Using Fourier's Law, the thermal properties of the film of interest can be derived.
	
In the 1990s, Cahill derived a method to determine the temperature fluctuation using a lock-in amplifier to extract a voltage signal with three times of the power source oscillating frequency\cite{cahill1990thermal}. A Wheatstone bridge is applied to attenuate the first harmonic signal without affecting the $3\omega$ component. Because other resistors of the Wheatstone bridge are placed in a stable environment temperature, the metal line on the top of the thin film, which is placed in a vacuum chamber with a radiation shield, can be considered as the only element in the circuit that is sensitive to the temperature fluctuation, and therefore is the only source that generates the $3\omega$ voltage\cite{dames2013measuring}. By extracting this signal, the temperature fluctuation of the metal line and the cross-plane thermal conductivity of the thin film can be derived as well as the effective cross-plane thermal conductivity of the thin film and the substrate. Hence the name 3$\omega$ method.
	
We define the temperature fluctuation of at the location of the probe, considering the voltage generated by the signal generator equal to $V = V_0 \sin(\omega t)$.
	 \begin{equation}
	\Delta T_{2\omega,rms}=2\frac{dT}{dR}\frac{V_{3\omega,rms}}{\beta V_0}
	\label{eq:temp_in_voltage}
	\end{equation}
	
In which, $\beta = \frac{R_h}{R_h+R_2}$ is the Wheatstone Bridge ratio. Since the temperature modulation will not be in the linear response regime, when the thermal penetration depth is relatively small, the cross-plane thermal conductivity with a Wheatstone Bridge applied is\cite{Daxi2018S},	 
	 \begin{equation}
	k_y=\frac{\beta^3 V_0^3}{4 \pi l R_{h0} V_{3\omega,rms} } \frac{dT}{dR} \int_0^{\infty} \frac{\sin^2(\zeta w)}{(\zeta w)^2(\zeta^2+\varrho^2)^{1/2}}d\zeta.
	\label{eq:cross_plane}
	\end{equation}	

\begin{center}	
\begin{figure}[ht!]
\includegraphics[scale=0.3]{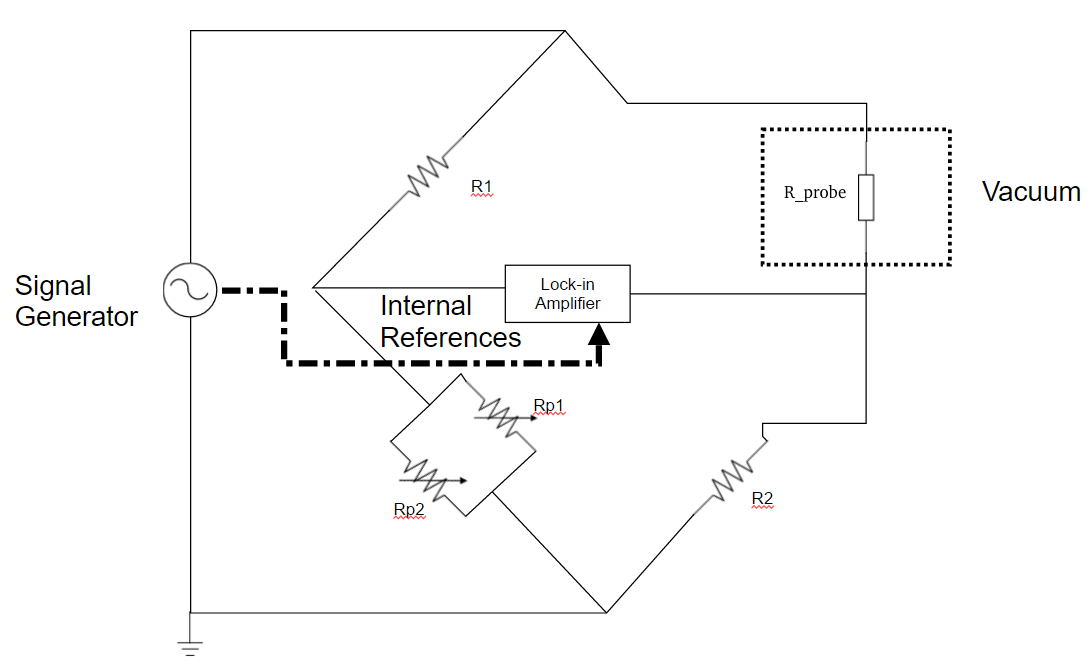}
\caption{Circuit diagram for the cross-plane thermal conductivity measurement}
\label{fig:circult_cross}
\end{figure}
\end{center}
	 
Eq.~\eqref{eq:cross_plane} modifies the effect of finite heater width with the application of a Wheatstone Bridge. Despite calling $k_y$ the cross-plane thermal conductivity, it is more appropriate to name this property as the effective thermal conductivity. The majority of the heat signal observed by the metal line is due to the cross-plane thermal conductivity, however the heat fluctuation contour reaches to the in-plane direction as well. Therefore $k_y$ is a combination of mostly the cross-plane thermal conductivity along with small portions of the in-plane thermal conductivity. For isotropic materials, this value is equal to the actual cross-plane thermal conductivity.
	
\begin{figure}[ht!]
	\begin{center}
	\includegraphics[scale=0.25]{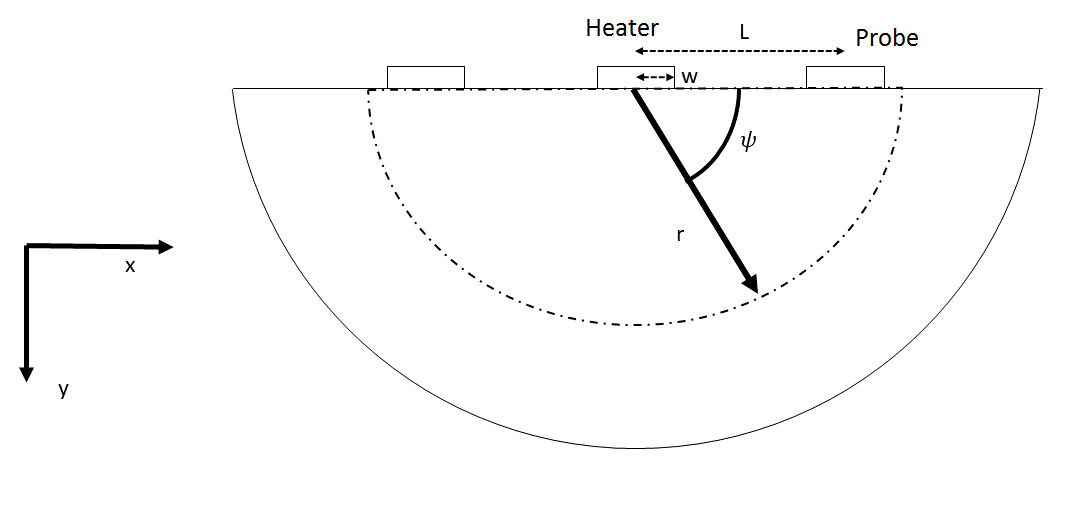}
	\caption{In-plane thermal conductivity measurement probe configuration}
	\label{fig:inplane}
	\end{center}
	\end{figure}

In the traditional 3$\omega$ method, the metal line is used as a heater and a probe, and this will cause a small temperature fluctuation $\Delta T_{2\omega, rms}$ at the location of the metal line. In order to solve the in-plane thermal conductivity, an extra metal line is introduced. The distance between two metal line is $L$. In the configuration displayed in Fig. \ref{fig:inplane}, if the voltage across the heater is much larger than the voltage across the probe, it would cause a large temperature fluctuation across the thin film. The probe then would be able to measure the temperature fluctuation caused by the heater. Combining with the temperature fluctuation caused by the probe itself $\Delta T_{2\omega, rms}$, the total temperature fluctuation $\Delta T_{probe, rms}$ is reflected by the third harmonic voltage across the probe. In mathematical term, the temperature fluctuation of the probe is,	
	 \begin{equation}
	\Delta T_{probe,rms}=\Delta T_{2\omega,rms}+\Delta T_{heater,rms}.
	\label{eq:temp_change_inplane}
	\end{equation}
	
The term $\Delta T_{heater}$ can be considered as a temperature fluctuation that is caused by the heater via in-plane thermal conductivity only. Therefore, using this temperature fluctuation, the in-plane thermal conductivity can be successfully obtained. The RMS temperature fluctuation measured at the probe due to the heater $\Delta T_{heater,rms}$ is,	
	\begin{equation}
	\Delta T_{heater,rms}=\frac{2}{\beta V_0}\frac{dT}{dR}(V_{probe,3\omega, rms}-V_{3\omega,rms}).
	\label{eq:V_3omega_rms}
	\end{equation}
	
The term $V_{3\omega,rms}$ is the RMS value of the 3$\omega$ voltage due to the temperature fluctuation of the probe itself (the 3$\omega$ signal for the traditional metal line configuration). After connecting the heater, the temperature fluctuation of the probe will change, hence resulting a new 3$\omega$ voltage $V_{probe,3\omega,rms}$. In order to understand the meaning of $\Delta T_{heater,rms}$ as well as the ability to solve the in-plane thermal conductivity, we start by analyzing the heat equation in two dimension, as the following,	
	\begin{equation}
	\rho C \frac{\partial \theta}{\partial t}=k_x\frac{\partial^2 \theta}{\partial x^2}+k_y\frac{\partial^2 \theta}{\partial y^2}.\\
	\end{equation}	 
	
For simplicity, we use complex time-dependent temperature fluctuation $\theta(x,y,t)$. The real time-dependent temperature fluctuation is $\Delta T=\Re(\theta)$. If we define the anisotropic ratio $r_k$ as the ratio of the in-plane thermal conductivity and the cross-plane thermal conductivity $r_k=\frac{k_y}{k_x}$, dividing both side of the equation by $k_x$, we get,	
	\begin{equation}
	\frac{1}{\alpha_x}\frac{\partial \theta}{\partial t}=\frac{\partial^2 \theta}{\partial x^2}+r_k \frac{\partial^2 \theta}{\partial y^2}.\\
	\label{eq:heat_2D_1}
	\end{equation}	 
	
	\begin{figure}[ht!]
	\begin{center}
	\includegraphics[scale=0.35]{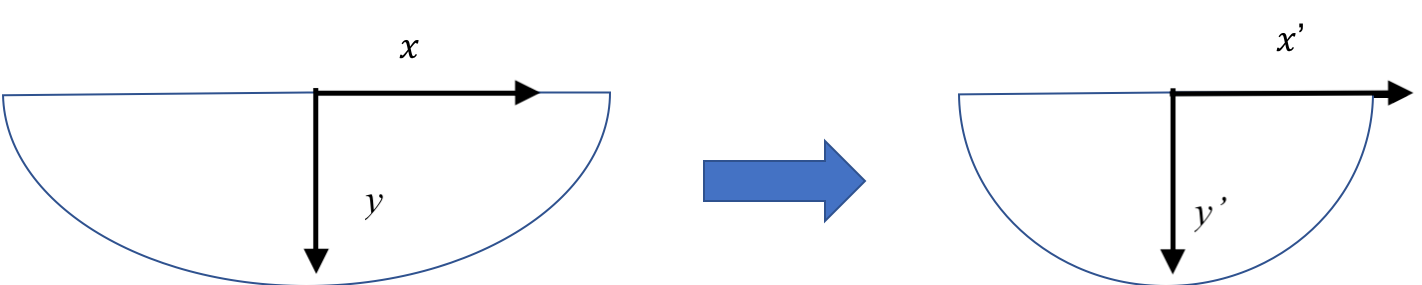}
	\caption{Normalization of the heat fluctuation contour for anisotropic material}
	\end{center}
	\end{figure}
	
Defining $x'=x$ and $y'=\sqrt{r_k}y$, we can normalize the heat fluctuation contour from elliptical to circular and will transform the coordinate system from $x,y$ to $x',y'$. According to Ref. \cite{carslaw1959conduction}, the solution has the format of $\theta=\Theta\exp(i2\omega t)$. Here $\Theta=\Theta(x',y')=\Theta(x,\sqrt{r_k}y)$. As a result the equation \eqref{eq:heat_2D_1} will reduce to,
	\begin{equation}
	\frac{\partial^2 \Theta}{\partial x'^2}+\frac{\partial^2 \Theta}{\partial y'^2}=\frac{i 2 \omega}{\alpha_x}\frac{\partial T}{\partial t}=\varrho_x'^2 \frac{\partial \Theta}{\partial t}.\\
	\end{equation}
	
The term $\varrho_x'=\sqrt{\frac{i 2 \omega}{\alpha_x}}$, is the wave number of the thermal wave in the in-plane direction. Compared with ref.\cite{cahill1990thermal}, the definition of $\varrho_x'$ is similar to the wave number of the thermal wave in the cross-plane direction, with a minor difference in the denominator. The magnitude of $1/\varrho_x'$ can be named as the thermal penetration length in the $x'$ direction $\delta_{p,x}$, indicating the amplitude of the temperature change being reduced by $90\%$ relative to the heater in the horizontal direction\cite{bergman2011fundamentals}. It is the minimum distance, for which the temperature fluctuation can be neglected . The film can be considered as semi-infinite if the thermal penetration depth in the $y$ direction is much less than the film thickness.

Now transferring $x',y'$ from Cartesian coordinate to cylindrical coordinate with $r',\psi'$, where $r'=\sqrt{x'^2+y'^2}$ and $\psi=\tan^{-1}(\frac{y'}{x'})$ we have,
	\begin{equation}
	\frac{\partial^2 \Theta}{\partial r'^2}+\frac{1}{r'}\frac{\partial \Theta}{\partial r'}+\frac{1}{r'^2}\frac{\partial^2 \Theta}{\partial \psi'^2}-\varrho_x^2 \Theta=0.
	\end{equation}
	
Separating the solution of $r'$ and $\psi'$, that is assumed to be $\Theta=\Theta_{r'}\Theta_{\psi'}$, then we can get the series solution.
	\begin{equation}
	\begin{split}
		\Theta(r',\psi')=\sum_{n=0,1,2...}&[\left( a_n I_n(\varrho_x r')+b_n K_n(\varrho_x r')\right) \\
		~~~&\cdot\left(c_n\exp(-i n \psi')+d_n \exp(i n \psi') \right)].
	\label{eq:heat_2d_primary_solution}
	\end{split}
	\end{equation}
	
In this equation, the terms $a_n,b_n,c_n$ and $d_n$ are the expansion coefficients which are to be determined from the boundary conditions. $I_n$ and $K_n$ are the modified first kind and second kind of Bessel functions. However before applying boundary conditions, this equation can be simplified. First, $n$ must be even, non-negative integers. We assumed that $k_x$ at $\psi=0$ is equal to $k_x$ at $\psi=\pi$ and therefore the temperature profile is symmetric from the heater, this indicates that $n$ is an even number. Moreover we can neglect the negative integers because the solutions of negative integers for the Bessel functions are not linearly independent. The temperature will decay to $T_{\infty}$ as $r'$ increases, therefore the constant in front of the first kind modified Bessel function must be equal to zero, since the result of first kind modified Bessel function increases to infinity as $r'$ increases. Finally, if we let $B_n=b_n c_n$ and $D_n=d_n/c_n$, Eq.\eqref{eq:heat_2d_primary_solution} can be simplified to,	
	\begin{equation}
	\begin{split}
	\Theta(r',\psi')=\sum_{n=0,2,4...}&\left(B_n K_n(\varrho_x r')\right)\cdot(\exp(-i~n \psi') \\ \nonumber &+D_n \exp(i~n \psi')).
	\end{split}
	\label{eq:heat_2d_solution_before_transform}
	\end{equation}
	
Now by transforming this equation back to the original coordinate system, we get,

	\begin{equation}
	\begin{split}
	\Theta&(r,\psi) =\sum_{n=0,2,4...}\left[B_n K_n\left(\varrho_x r \sqrt{1+(\frac{1}{r_k}-1)\sin^2{\psi}}\right)\right]
	\cdot \lbrace \exp\left[-i~n \cos^{-1}\left(\cos{\psi}/\sqrt{\cos^2{\psi}+\frac{\sin^2{\psi}}{r_k}}\right)\right]\\
	&~ +D_n \exp\left[i~n \cos^{-1}\left(\cos{\psi}/\sqrt{\cos^2{\psi}+\frac{\sin^2{\psi}}{r_k}}\right)\right] \rbrace.
	\label{eq:heat_2d_solution_after_transform}
	\end{split}
	\end{equation}
	
The goal now is to solve the constants $B_n$ and $D_n$ based on boundary conditions. In order to do so, we consider a semi-cylindrical region $\Gamma$ with radius $r_0$ around a thin and narrow heater. This region $\Gamma$ is bounded by two parts, the flat surface region on the top, referred to as $\Omega'$, and the half cylindrical surface region on the bottom, referred to as $\Omega$. The energy balance of this volume $\Gamma$ is,
	\begin{equation}
	E_{in}-E_{out}+E_{generation}=E_{storage}.
	\label{eq:energy_balance}
	\end{equation}	
	\begin{figure}
	\begin{center}
	\includegraphics[scale=0.25]{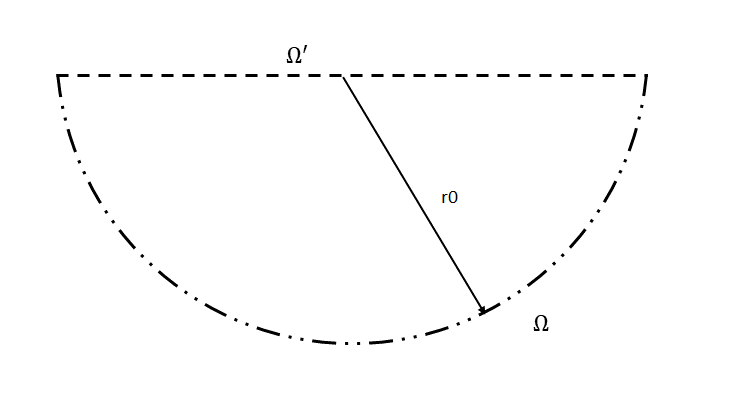}
	\caption{Closed semi-infinity small cylinderical volume illustration}
	\label{fig:semi-infinite}
	\end{center}
	\end{figure}

Since there is no heat generation inside of the film, the generation term can be ignored. Additionally, the energy entering the system $E_{in}$ will be the power from the heater, 
	
	$$E_{in}=\frac{\beta^2 V_0^2 \exp(i2\omega t)}{R_{h0}}.$$
	
And, the energy leaving the system $E_{out}$ will be the heat flux multiplied by the surface area,
	
	$$E_{out}=\overrightarrow{q}A=-\exp(i2\omega t) \int_{\Omega} \overrightarrow{r}\cdot(\overrightarrow{k} \cdot \nabla T) d\Omega.$$
	
The outward normal unit vector of the cylindrical surface $\Omega$ is,
	
	$$\overrightarrow{r}=\overrightarrow{i}\cos(\psi)+\overrightarrow{j}\sin(\psi).$$
	
The term $\overrightarrow{i}, \overrightarrow{j}$ are unit vector in $x,y$ direction. The energy stored within the system $E_{st}$ is,
	
	$$E_{st}=\rho c V \frac{dT}{dt}=l \rho c \exp(i2\omega t)\int_{r=0}^{b}\int_{\psi=0}^{\pi} r i 2 \omega \Theta~ dr d\psi. $$
	
Now plug back into Eq.\eqref{eq:energy_balance}, the time dependent term $\exp(i2\omega t)$ can be dropped and we have,
	\begin{equation}
	\begin{split}
	\frac{\beta^2 V_0^2}{R_{h0}}+ \int_{\Omega} \overrightarrow{r}&\cdot(\overrightarrow{k} \cdot \nabla \Theta) d\Omega=\\
	&l \rho c (i 2 \omega)\int_{r=0}^{r_0}\int_{\psi=0}^{\pi}  \Theta r~ dr~d\psi,
	\label{eq:whatever}
	\end{split}
	\end{equation}
	
in which the term $l$ represents the length of the metal line. Expanding the surface integral we get,
	\begin{equation}
	\begin{split}
	\int_{\Omega} \overrightarrow{r}\cdot(\overrightarrow{k} \cdot \nabla \Theta) d\Omega=\int_{\Omega} 
	\left[
	\begin{array}{cc}
	\cos\psi &\sin\psi
	\end{array}
	\right]
	\left[
	\begin{array}{c}
	\overrightarrow{i}\\
	\overrightarrow{j}
	\end{array}
	\right] \\
	\cdot\left(
	\left[
	\begin{array}{cc}
	k_x & 0\\
	0 & r_k k_x
	\end{array}
	\right]
	\left[
	\begin{array}{cc}
	\frac{\partial \Theta}{\partial x}& 0\\
	0& \frac{\partial \Theta}{\partial y}
	\end{array}
	\right]
	\left[
	\begin{array}{c}
	\overrightarrow{i}\\
	\overrightarrow{j}
	\end{array}
	\right]
	\right) d\Omega.
	\end{split}
	\label{eq:half_result}
	\end{equation}

This equation is in Cartesian coordinates. Unfortunately the expression for $\Theta$ is in cylindrical coordinate. In order to transfer the result to cylindrical coordinates, we have to use a transfer matrix on the unit vector. We can also apply the chain rule on the derivative. The transfer matrix is described as follows,
	\begin{equation}
	\left[
	\begin{array}{c}
	\overrightarrow{i}\\
	\overrightarrow{j}
	\end{array}
	\right]=
	\left[
	\begin{array}{cc}
	\cos\psi &-\sin\psi \\
	\sin\psi &\cos\psi 
	\end{array}
	\right]
	\left[
	\begin{array}{c}
	\overrightarrow{r}\\
	\overrightarrow{\psi}
	\end{array}
	\right].
	\label{eq:transf}
	\end{equation}
	
Therefore Eq.\eqref{eq:half_result} becomes,
	\begin{equation}
	\begin{split}
\int_{\Omega} \overrightarrow{r} \cdot(\overrightarrow{k} \cdot \nabla \Theta) d\Omega=
l\int_{\psi=0}^{\pi} \left(r k_x\frac{\partial \Theta}{\partial r}(\cos^2\psi +r_k\sin^2\psi) \right)\\
+\left(k_x\frac{\partial \Theta}{\partial \psi}\sin\psi\cos\psi(1+r_k)\right)d\psi,
	\end{split}
	\label{eq:middle step}
	\end{equation}

in which the symbol $\Theta$ is defined by the Eq.\eqref{eq:heat_2d_solution_after_transform}. In general, the derivative of second kind of modified Bessel functions equals to $\frac{\partial K_n(z)}{\partial z}=\frac{1}{2}(-K_{n-1}(z)-K_{n+1}(z))$ for positive integer $n$ such that $n\geq1$. Therefore, the terms $\frac{\partial \Theta}{\partial r}$ and $\frac{\partial \Theta}{\partial \psi}$ will be series of multiplications of second kind of modified Bessel function. Now we assume that the radius $r_0$ is approaching to $0$. Knowing that $K_n(z) \propto 1/z ^n $ (where $n \geq 1$), $|z|\ll 1$, we can take the limit of Eq.\eqref{eq:middle step} when $r_0\rightarrow 0$, 
	 \begin{equation}
	 \begin{split}
	 \lim_{r_0\rightarrow 0}\int_{\Omega} \overrightarrow{r} \cdot(\overrightarrow{k} \cdot \nabla \Theta) d\Omega=\\
	 \lim_{r_0\rightarrow 0} l\int_{\psi=0}^{\pi} \left(r k_x\frac{\partial \Theta(r,\psi)}{\partial r}(\cos^2\psi +r_k\sin^2\psi)\right)\\
	 + \left( k_x\frac{\partial \Theta(r,\psi)}{\partial \psi}\sin\psi\cos\psi(1+r_k)d\psi \right).
	 \end{split}
	 \label{eq:blowup}
	 \end{equation}
	 
The right hand side of Eq.\eqref{eq:middle step} will diverge for all $n\geq 0$. Therefore in order to receive meaningful results, we must have $B_n=0$ for all $n\geq 1$ leaving only $B_0$ undetermined. For all $n\geq 1$, $B_n$ are multiplied with another function that contains a series of constants $D_n$. $D_n$ is irrelevant in this case. If we assume for $n\geq 1$, $D_n=0$ and re-write the solutions for $\Theta$ we get,
	\begin{equation}
	\begin{split}
	\Theta(r,\psi)& =B_0 K_0\left(\varrho_x r \sqrt{1+(\frac{1}{r_k}-1)\sin^2{\psi}}\right)\\
	&~~~~\cdot(\exp(0)+D_0 \exp(0))\\
	&=A_0 K_0\left(\varrho_x r \sqrt{1+(\frac{1}{r_k}-1)\sin^2{\psi}}\right),
	\end{split}
	\label{eq:heat_2d_solution_final}
	\end{equation}
	
where $A_0=B_0(1+D_0)$. This represents the temperature profile for the anisotropic case. This equation can be easily checked by setting $r_k=1$, which indicates the isotropy of the film. Then $\Theta(r,\psi)=A_0 K_0\left(\varrho_x r\right)$. The result agrees with result from ref \cite{cahill1990thermal}.
		
For the zeroth order modified Bessel function of second kind, this relation is valid for any constant $a\neq 0$,
	 \begin{equation}
	 \lim_{b\rightarrow 0}\int_{r=0}^{b}K_0(ar)rdr=0
	 \end{equation}
	 
If we take the limit on the right hand side of Eq.\eqref{eq:whatever}, it will cause the volume integral to vanish. It agrees with the assumption that an infinitely small volume does not store energy. This result brings us to the energy balance equation for $r\rightarrow 0$,
\begin{equation}
	\frac{\beta^2 V_0^2}{2R_{h0}}=- \int_{\Omega} \overrightarrow{r}\cdot(\overrightarrow{k} \cdot \nabla \Theta) d\Omega.
	\label{eq:energy_balance_simple}
	\end{equation}

The Puiseux series expansion for modified Bessel function of the second kind is,
	\begin{equation}
	K_1(x)_{x=0}=\frac{1}{x}+O(x^2).
	\end{equation}	
	
The derivative of $\Theta$ with respect of $r$ evaluated at $r_0$ where $r_0\rightarrow 0$ is,
	\begin{equation}
	\begin{split}
	\lim_{r_0\rightarrow 0}\frac{\partial \Theta}{\partial r}|_{r=r_0}
	&\approx\lim_{r_0\rightarrow 0}-\frac{A_0}{r_0}.
	\end{split}
	\end{equation}
	
The derivative of $\Theta$ with respect of $\psi$ is,
	\begin{equation}
	\begin{split}
	\lim_{r_0\rightarrow 0}\frac{\partial \Theta}{\partial \psi}|_{r=r_0}
	&\approx\lim_{r_0 \rightarrow 0}\frac{A_0\sin\psi \cos\psi}{1+(\frac{1}{r_k}-1)\sin^2{\psi}}.
	\end{split}
	\end{equation}
	 
Combining Eqs.\eqref{eq:blowup} and \eqref{eq:energy_balance_simple} with taking the limit of $r \rightarrow r_0$, we have,
	\begin{equation}
	\begin{split}
	\frac{\beta^2 V_0^2}{2 R_{h0}} &=-\lim_{r_0\rightarrow 0} l\int_{\psi=0}^{\pi}(r k_x\frac{\partial \Theta(r,\psi)}{\partial r}(\cos^2\psi +r_k\sin^2\psi)\\
	&~~~~~~~~~ +k_x\frac{\partial \Theta(r,\psi)}{\partial \psi}\sin\psi\cos\psi(1+r_k))d\psi\\
	&= l k_x A_0(\frac{\pi}{2}+r_k\frac{\pi}{2}).
	\end{split}
	\label{eq:66}
	\end{equation}
	
Therefore solving for $A_0$, we get,
	\begin{equation}
	A_0=\frac{\beta^2 V_0^2}{\pi R_{h0}l k_x(1+r_k)}.
	\end{equation}	
	
Note that the assumption of $r_0 \rightarrow 0$ is to create a semi-infinite small volume. For such small volume, the energy stored within the volume is approximately equal to zero. It is a valid assumption because during the actual measurement, the energy storage will be reflected as the temperature rises of the film/substrate system. Hence we use RMS voltage and temperature in the following calculation.
	
To check for the isotropic case, we let $r_k=1$, the result is  $A_0=\frac{\beta^2 V_0^2}{2 l k_x \pi R_{h0}}$, which agrees with ref.\cite{cahill1990thermal} 
	
Plug $A_0$ back into Eq.\eqref{eq:heat_2d_solution_final}, we have,
	\begin{equation}
	\begin{split}
	\Theta(r,\psi)=&\frac{\beta^2 V_0^2}{\pi R_{h0}l k_x(1+r_k)}\\
	&~~~~~~~~K_0\left(\varrho_x r \sqrt{1+(\frac{1}{r_k}-1)\sin^2{\psi}}\right).
	\label{eq:final_yeah}
	\end{split}
	\end{equation}
	
Eq.\eqref{eq:final_yeah} defines the temperature fluctuation at location $(r,\psi)$ due to the heater at $r=0$. This equation can be further simplified by converting the coordinate system back to Cartesian coordinates, which results the following equation,
	\begin{equation}
	\Theta(x,y)=\frac{\beta^2 V_0^2}{\pi R_{h0}l k_x(1+r_k)}K_0\left(\varrho_x \sqrt{x^2+\frac{1}{r_k} y^2}\right).
	\label{eq:final_yeah2}
	\end{equation}
	
Now consider Fig.\ref{fig:inplane}, at location $x= L, y=0$, the temperature fluctuation is,
	\begin{equation}
	\begin{split}
	\Delta T_{heater,rms}(L,0)&=\frac{\beta^2 V_0^2}{2 \pi R_{h0}l k_x(1+r_k)}\Re (K_0\left(\varrho_x L \right))\\
	&=\frac{2}{\beta V_0} \frac{dT}{dR}(V_{probe,3\omega, rms}-V_{3\omega,rms})
	\end{split}
	\end{equation}
	
now, solving for $k_x$, we find,
	\begin{equation}
	\begin{split}
	k_x&= \Re (K_0(\varrho_x L))\frac{dR}{dT} \\
	&~~~~~~~ \cdot \frac{\beta^3 V_0^3}{4 w R_{h0} l \pi (V_{probe,3\omega, rms}-V_{3\omega,rms})}-k_y,
	\label{eq:in_plane_k}
	\end{split}
	\end{equation}
	
in which the quantities are either known or measurable. 
\section{SAMPLE FABRICATION AND PREPARATION}
\subsection{Sample Fabrication}
In order to test the accuracy of Eq.\eqref{eq:in_plane_k}, the measurement can be performed on an isotropic material (such as SiN, metal or SiO$_2$). The in-plane thermal conductivity is expected to be the same as the cross-plane thermal conductivity. Then further measurement can be used to observe the in-plane thermal conductivity of an anisotropic material (such as graphite, h-BN). In this experiment we will validate this method on isotropic film materials. We purchased a 100nm pre-growth LPCVD silicon nitride film on $500\mu$m of $<100>$ type silicon wafers. We have also purchased a $500\mu$m of $<100>$ type silicon wafers for sputtering boron nitride thin films. The hexagonal boron nitride target is from Kurt J. Lesker, with a $99.5\%$ purity, 2" diameter and 0.125" thickness.

We were able to deposit a 64 nm thick amorphous boron nitride on top of $500\mu$m silicon wafer with a DC magnetron sputtering system. Although the system has a digital reader for estimating the thickness of the film based on the density of the material, it is necessary to use scanning electron microscope (SEM) to confirm the thickness of the film after deposition is finished. In order to calculate the effective cross-plane thermal conductivity of the film accurately, the thermal penetration depth needs to be much less than the thickness of the film. A poor estimate of the film thickness could result in the measurement of the effective cross-plane thermal conductivity of the substrate and the film instead of the thin film only.

\begin{figure}[ht!]
\begin{center}
\includegraphics[scale=0.3]{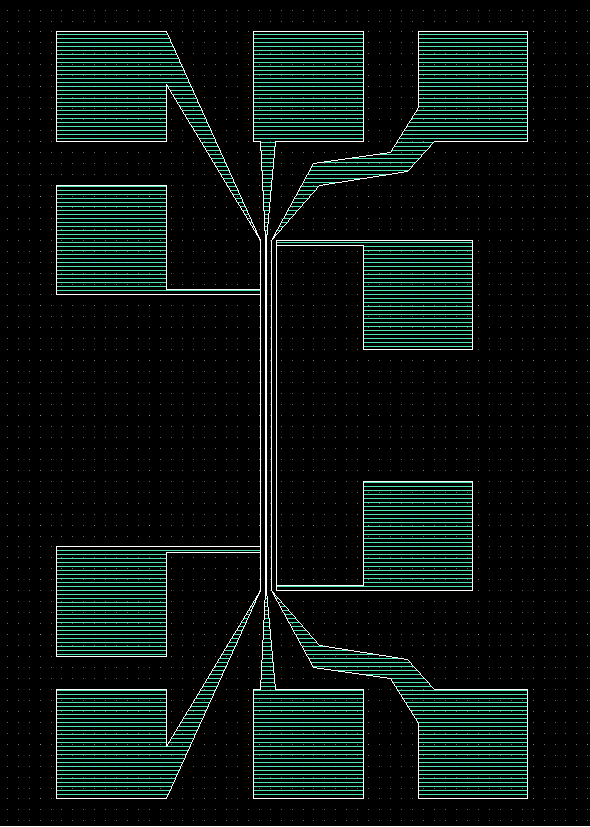}
\caption{Pattern design for the in-plane thermal conductivity measurement}
\label{fig:new_design}
\end{center}
\end{figure}

After fabricating the samples, the next step is to perform photolithography for metal line deposition. The pattern of the photolithography mask for the in-plane thermal conductivity measurement is displayed in Fig.\ref{fig:new_design}. In addition to the traditional four point probe design that can be used to accurately measure the resistances of the metal line, three extra metal lines with the same length and width, are placed next to the four point probe with equal distance between. The advantage of this design is that each individual metal line can be used for measuring the effective cross-plane thermal conductivity ($k_y$) whereas any two parallel metal lines (regardless whether they are next to each other or not) can be used to measure the in-plane thermal conductivity. Therefore by repeating the in-plane thermal conductivity measurement as well as the effective cross-plane thermal conductivity measurement at different locations, the overall accuracy can be improved.

The accuracy of the term $dR/dT$ will heavily influence the measured accuracy of the thermal properties. Therefore the material of the metal lines must have a large temperature coefficient of the resistance to create a measurable signal. In general, metals such as gold, aluminum, silver, and platinum are preferred because these materials have a good response to the temperature change at lower temperatures ($\sim$ 30 K) and higher temperatures ($\sim$ 750K)\cite{dames2013measuring}. In this experiment, platinum is deposited on the top of the film using a DC magnetron sputtering system. The platinum target was purchased from Kurt J. Lesker, with a purity of 99.9$\%$, 2" diameter and 0.0625" thickness. After the deposition step deposition, lift off and general cleaning, the length of each metal line as well as the distance between the metal lines, is measured using SEM.
\subsection{Measurement Procedure}
Before measuring the cross-plane or the in-plane thermal conductivity, one key parameter that needs to be measured is $dR/dT$, which is the temperature coefficient with respect of resistance of the metal lines. In Fig.\ref{fig:new_design} the probe on the left is prepared for the four-point probe measurement. After placing the sample inside of the vacuum chamber and adjusting the chamber's temperature using a temperature controller, the resistance of the metal line can be accurately measured over a wide temperature range. The term $dR/dT$, can be calculated by finding the derivative of curve-fitted equation of resistance with respect of temperature.

\begin{figure}[ht!]
\begin{center}
\includegraphics[scale=0.28]{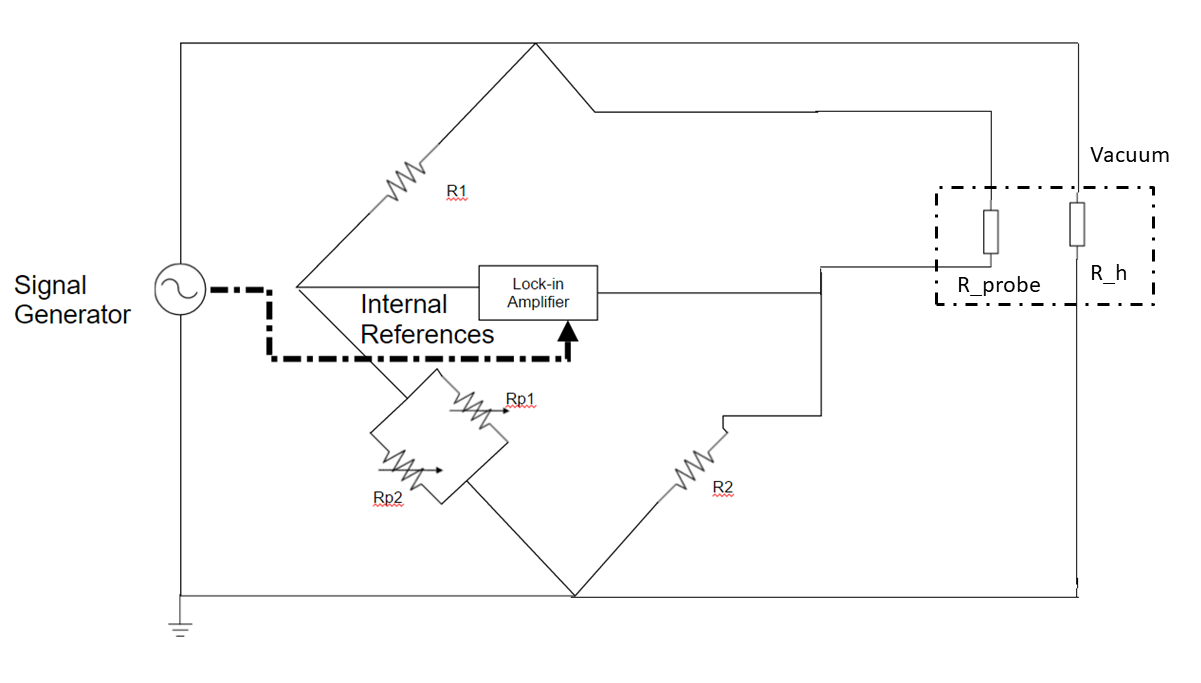}
\caption{Circuit configuration for the in-plane thermal conductivity measurement}
\label{fig:circuit_inplane}
\end{center}
\end{figure}

Comparing with the cross-plane thermal conductivity measurement which only uses one metal line as a probe and a heater, measuring the in-plane thermal conductivity requires separate probe and heater. Fig.\ref{fig:circuit_inplane} represent the circuit configuration for measuring the in-plane thermal conductivity: the heater is placed outside of the Wheatstone Bridge so that the voltage across the heater is equal to the voltage generated by the signal generator, and is in phase with the probe. The resistance ratio $\beta$ is required to be less than 0.1 for converting a voltage power supply into a voltage supply, therefore, the voltage across the probe is much less than the voltage across the heater.

The steps for measuring the in-plane thermal conductivity are described as follow: First of all, connect only the probe into the circuit, adjust the desired measuring temperature, and wait until the temperature reaches stability to balance the Wheatstone Bridge. Secondly, collect the 3$\omega$ voltage, $V_{3 \omega, rms}$, from the probe, and calculate the cross-plane thermal conductivity of the thin film $k_y$. Thirdly, connect the heater into the circuit. Because the voltage across the heater is much larger than the voltage across the probe, the temperature fluctuation generated by the heater will spread across the thin film, and will be captured by the heater. If the observation of the third harmonic voltage continues, a large voltage "jump" will occur when the heater is connected to the circuit. Last of all, wait until the voltage reaches stability to collect the new 3$\omega$ voltage, $V_{probe,3\omega,rms}$, and use Eq. \eqref{eq:in_plane_k} to calculate the in-plane thermal conductivity.

\begin{figure}[ht!]
\begin{center}
\includegraphics[scale=.5]{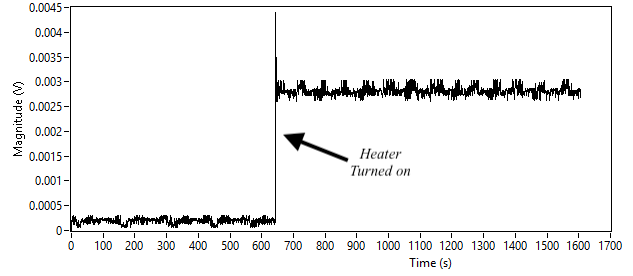}
\caption{The sudden increase of the third harmonic voltage}
\label{fig:3_V_I}
\end{center}
\end{figure}

\subsection{Results}
\begin{figure}[t!]
\begin{center}
\includegraphics[scale=0.35]{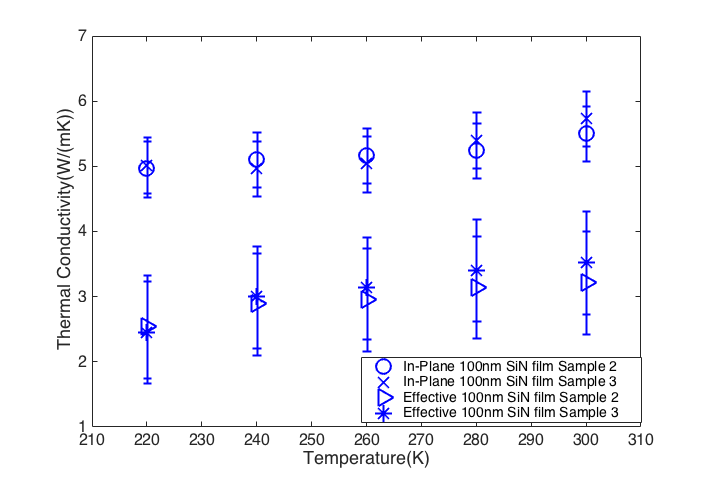}
\caption{The thermal conductivity of the 100 nm silicon nitride thin film}
\label{fig:in_plane_thermal_sin}
\end{center}
\end{figure}
Fig. \ref{fig:in_plane_thermal_sin} shows the cross-plane and in-plane thermal conductivities for the 100nm LPCVD silicon nitride film from $\sim 230$ K to 300 K. The thermal diffusivity of silicon nitride for both the in-plane and the cross-plane directions are $1.0 \times 10^{-6} m^2s^{-1}$, and the driving frequency for the cross-plane thermal conductivity are 20 MHz and 22 MHz. We used two different frequencies to increase accuracy since under such frequencies, the thermal penetration depth will be under 50 nm, which are less than the film's thickness. The slope method is not applicable because the frequency is too high. The driving frequency for the in-plane thermal conductivity measurement is 100 Hz. This frequency will maintain the thermal penetration length larger than the distance between the probe and the heater, which is 100 $\mu$m in this case.

\begin{figure}[ht!]
\begin{center}
\includegraphics[scale=0.375]{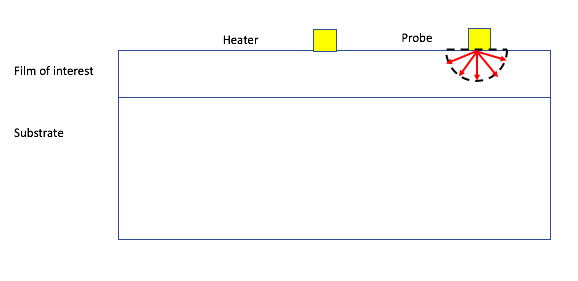}
\includegraphics[scale=0.38]{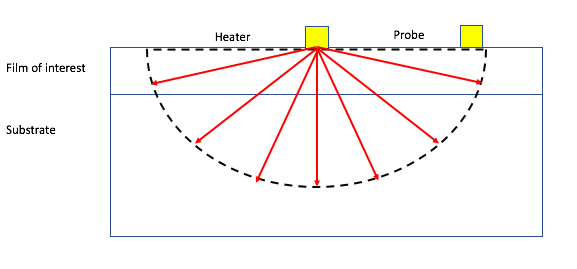}\\
\includegraphics[scale=.3]{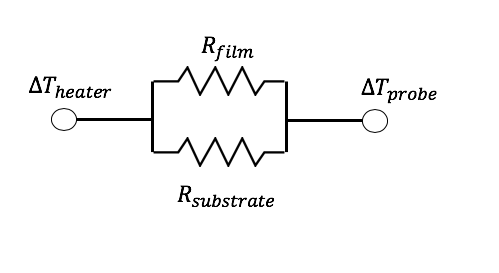}
\caption{Thermal fluctuation signal illustration}
\label{fig:thermal circuit}
\end{center}
\end{figure}

\begin{figure}[ht!]
\begin{center}
\includegraphics[scale=0.35]{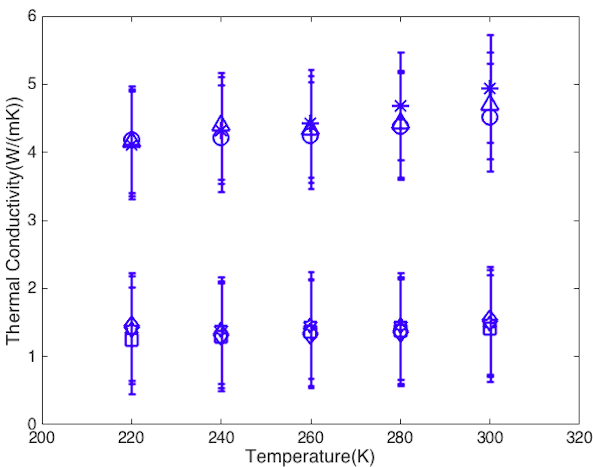}\\
\includegraphics[scale=0.3]{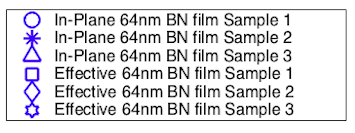}
\caption{The thermal conductivity of the 64 nm amorphous boron nitride thin film}
\label{fig:cross_BN}
\end{center}
\end{figure}

Here we observed anisotropic behavior in an isotropic material. This is due to the probe layout and the composite structure of film on substrate. The distance between the probe and the heater is much larger than the film thickness. Therefore, in order to create a thermal fluctuation that can be captured by the probe, the thermal fluctuation caused by the heater will penetrate through the film and into the substrate. Therefore, despite the film being isotropic, the composite structure introduces anisotropy into the final results. From a thermal circuit's perspective, which illustrated in Fig.\ref{fig:thermal circuit}, two thermal resistors are in parallel. Therefore the effective thermal resistance is reduced, causing the in-plane thermal conductivity to be larger than the cross-plane thermal conductivity.
In order to test this speculation, we performed the same measurement for the 64 nm amorphous boron nitride thin film sample. Fig.\ref{fig:cross_BN} shows the cross-plane and the in-plane thermal conductivity for the 64 nm sputtered amorphous boron nitride film from 230 K to 300 K. The thermal diffusivity of boron nitride for both the in-plane and the cross-plane direction are $1.0 \times 10^{-7} m^2s^{-1}$, and the driving frequency for the cross-plane thermal conductivity are 26 MHz and 27 MHz. Such frequencies will maintain the thermal penetration depth under 43 nm. The driving frequency for the in-plane thermal conductivity measurement is 100 Hz.

The results of the boron nitride sample matches with the speculation, that is, we have observed anisotropy within an isotropic material, and the in-plane thermal conductivity is larger than the cross-plane thermal conductivity.
\section{CONCLUSION AND DISCUSSION}
Compared with other in-plane thermal conductivity measurements of thin film materials, the method presented in this study provides a simple transient solution for the direct measurement in-plane and cross-plane thermal conductivities of anisotropic materials. The simplicity of the method is based on the assumption that the thermal fluctuation above the thermal penetration depth (or length) can be ignored.

However the accuracy of estimating the thermal penetration
depth will impact the results. Ideally, when the thermal penetration depth is greater than the film thickness, the thermal wave will penetrate the film and enter the substrate. In reality, there is a boundary thermal resistance between the film and the substrate. This could change overall thermal resistance in general. A more accurate method to determine thermal the boundary resistance requires further study.

This work demonstrate a technique that allows for the direct
measurement of the in-plane and cross plane thermal conductivity
of anisotropic materials. This will accumulate the study
of anisotropic material for heat spreading in nanoelectronic
devices.

\bibliography{Arxiv}

\end{document}